\documentclass[12pt,a4paper]{article}
\usepackage{jheppub}

\newcommand{\de}{\mathrm{d}}

\newcommand{\I}{\mathrm{i}}

\newcommand{\cD}{\mathcal{D}}
\newcommand{\cF}{\mathcal{F}}
\newcommand{\cI}{\mathcal{I}}

\newcommand{\cM}{\mathcal{M}}

\newcommand{\cE}{\mathcal{E}}

\newcommand{\cR}{\mathcal{R}}

\newcommand{\cO}{\mathcal{O}}
\newcommand{\cH}{\mathcal{H}}

\newcommand{\IR}{\mathbb{R}}

\newcommand{\IZ}{\mathbb{Z}}

\newcommand{\zetastar}{\zeta^\star}

\def\bea{\begin{eqnarray}}
\def\eea{\end{eqnarray}}
\def\be{\begin{equation}}
\def\ee{\end{equation}}
\def\ba{\begin{align}}
\def\ea{\end{align}}
\def\bse{\begin{subequations}}
\def\ese{\end{subequations}}
\def\Im{\,{\rm Im}\,}
\def\Re{\,{\rm Re}\,}

\def\half{\frac{1}{2}}

\def\RN{{\rm R.N.}}

\def\rr{\rho_d}

\title{Infrared divergences and harmonic anomalies in
the two-loop superstring effective action}

\preprint{\hfill\parbox[t]{5cm}{CERN-PH-TH-2015-240\\QMUL-PH-15-17}}

\author[a,b,c]{Boris Pioline,} 

\author[d]{Rodolfo Russo}
 
 \affiliation[a]{CERN PH-TH,
Case C01600, CERN, CH-1211 Geneva 23, Switzerland}

\affiliation[b]{Sorbonne Universit\'es, UPMC Universit\'e Paris 6, UMR 7589, F-75005 Paris, France}
 
\affiliation[c]{ Laboratoire de Physique Th\'eorique et Hautes
Energies, CNRS UMR 7589, \\
Universit\'e Pierre et Marie Curie,
4 place Jussieu, 75252 Paris cedex 05, France} 

\affiliation[d]{Centre for Research in String Theory, School of Physics and Astronomy,\\
Queen Mary University of London, Mile End Road, London, E1 4NS, UK}

\emailAdd{boris.pioline@cern.ch}
\emailAdd{r.russo@qmul.ac.uk}
 
 \abstract{We analyze the pertubative contributions to the $D^4\cR^4$ and $D^6\cR^4$ couplings in the  low-energy effective action of type II string theory compactified on a torus $T^d$, with particular emphasis on two-loop corrections. In general, it is necessary to introduce an infrared cut-off $\Lambda$  to separate local interactions from non-local effects due to the exchange of massless states. We identify the degenerations of the genus-two Riemann surface which are responsible for power-like dependence on $\Lambda$, and give an explicit prescription for extracting the $\Lambda$-independent effective couplings. These renormalized couplings are then shown to be eigenmodes of the Laplace operator with respect to the torus moduli, up to computable anomalous source terms arising in the presence of logarithmic divergences, in precise agreement with predictions from U-duality. Our results for the two-loop $D^6\cR^4$ contribution also probe essential properties of the Kawazumi-Zhang invariant. }

\begin{document}
\maketitle

\section{Introduction}

Supersymmetry and duality provide strong constraints on the possible higher derivative corrections to the low-energy effective action in flat type II string vacua with maximal supersymmetry. Combined with explicit scattering amplitude calculations at low order in string perturbation theory, these constraints sometimes completely determine the dependence of these couplings on all moduli, including the string coupling, through a suitable U-duality invariant automorphic function. Expanded at weak coupling, this function reveals, along with the few perturbative contributions which it was designed to reproduce, an infinite series of non-perturbative instanton effects, providing useful constraints on an eventual non-perturbative definition of string theory. 

This line of research has been carried through with great success for four-graviton couplings in type II string  compactified on a $d$-dimensional torus down to any dimension $D=10-d\geq 3$ \cite{Green:1997tv,Green:1997di,Kiritsis:1997em,Pioline:1997pu,Pioline:1998mn,Green:1998by,Obers:1999um,Green:1999pv,Basu:2007ru,Green:2010wi,Pioline:2010kb,Green:2010kv,Green:2011vz,Basu:2011he,Bossard:2014lra,Bossard:2014aea}. The leading term in the low-energy expansion corresponds to the Einstein-Hilbert term $\cR$ and its supersymmetric completion, which is protected from quantum corrections. Subleading terms correspond to terms schematically of the form $\cE^{(d)}_{(m,m)} D^{4m+6n} \cR^4$, where $D^{4m+6n} \cR^4$ denotes a specific combination of $4m+6n$ space-time derivatives and four powers of the Riemann tensor \cite{Green:1999pv}, and $\cE^{(d)}_{(m,m)}$ is a function on the symmetric moduli space $E_{d+1}/K_{d+1}$, invariant under the action of the U-duality group $E_{d+1}(\IZ)$ (here $E_{d+1}$
refers to the split real forms of the exceptional Lie groups $E_6, E_7, E_8$ for $d\geq 5$, or of the
classical Lie groups $A_1, A_1\times A_2, A_4,D_5$ for $d<5$). The coefficients 
$\cE^{(d)}_{(0,0)}$ and $\cE^{(d)}_{(1,0)}$ of the next-to-leading and
next-to-next-to-leading terms are known to be given by  suitable (residues of) Langlands-Eisenstein series for the U-duality group. This is consistent with the fact that supersymmetry requires these functions to be  eigenmodes of the Laplacian on the moduli space $E_{d+1}/K_{d+1}$ with a specific eigenvalue (up to anomalous terms for special values of the dimension $d$ where the local and non-local parts of the effective action mix) \cite{Green:2010wi,Green:2010sp} (see \cite{Bossard:2014lra,Bossard:2014aea,Wang:2015jna} for new perspectives on these supersymmetry constraints). Moreover,  the non-vanishing perturbative contributions (up to one-loop for $\cE^{(d)}_{(0,0)}$, and up to two-loop for $\cE^{(d)}_{(1,0)}$) are themselves known to be (residues of) Langlands-Eisenstein series for the T-duality 
group $SO(d,d,\IZ)$ \cite{Obers:1999um,Green:2010wi,Angelantonj:2011br,Pioline:2014bra}, consistently with the fact that the full non-perturbative couplings are  (residues of) Langlands-Eisenstein series for the U-duality group. 

In particular, the two-loop contribution to $D^4\cR^4$ is given by the modular integral
\be
\label{dfourrfour2}
\cE^{(d,2)}_{(1,0)}(G,B) =\frac{\pi}{2}\,  \RN \int_{\cF_2} \de\mu_2 \, \Gamma_{d,d,2}(\Omega;G,B)\,\ , 
\ee
where $\cF_2$ is the fundamental domain of the moduli space $\cM_2$ of compact Riemann surfaces
of genus 2, parametrized by the  period matrix $\Omega=\Omega_1+\I\Omega_2$, $\de\mu_2$ is the 
invariant measure  on $\cM_2$ normalized as in \cite{D'Hoker:2014gfa}\footnote{
$\de\mu_h = {(\det\Im\Omega)^{-h-1}} { \prod_{I\leq J} \I\, \de\Omega_{IJ} \wedge \de\bar\Omega_{IJ}}$}, and 
$\Gamma_{d,d,h}(\Omega;G,B)$ is the genus $h$ Narain lattice partition function  defined in \eqref{eq:Hf}, which depends on 
$\Omega$ and the metric $G_{ij}$ and Kalb-Ramond field $B_{ij}$ on the internal torus $T^d$.
The symbol $\RN$ stands for a renormalization prescription, which is necessary in dimension $d\geq 3$ due to infrared divergences (see below). By construction, the modular integral \eqref{dfourrfour2}
is an automorphic form on the Grassmannian $SO(d,d,\IR)/(SO(d)\times SO(d))$ parametrized by
$(G,B)$, invariant under T-duality. It is  proportional to the spinor Eisenstein series $E^{SO(d,d)}_{S,s=2}$ when $d>4$ (or to the sum $\hat E^{SO(d,d)}_{S,s=2}+ \hat E^{SO(d,d)}_{C,s=2}$ of the two regularized spinor Eisenstein series when $d\leq 4$) \cite{Obers:1999um}, and satisfies the Laplace  equation
\be
\left( \Delta_{SO(d,d)}  + d(d-3) \right)\, \cE_{(1,0)} ^{(d,2)}   =  24\zeta(2)\, \delta_{d,3}
+4 \cE_{(0,0)}^{(d,1)} \delta_{d,4}\ ,
\label{delsod4r4}
\ee
where $\cE_{(0,0)}^{(d,1)}$ is the one-loop contribution to the $\cR^4$ coupling,
\be
 \label{r4oneloopa}
\cE_{(0,0)} ^{(d,1)}(G,B)  = \pi\, \RN \int_{\cF_1} d \mu_1\, \Gamma_{d,d,1} (\tau; G,B)\ .
\ee
We shall refer to the anomalous terms appearing on the r.h.s. when $d=3$ or $d=4$ as `harmonic anomalies'. They follow from 
similar anomalous terms appearing in the U-duality invariant Laplace-type equation
for full $D^4\cR^4$ coupling $\cE_{(0,0)}^{(d)}$, which 
were determined in \cite{Pioline:2015yea} using general consistency requirements and confirmed in \cite{Bossard:2015oxa}. They can also be extracted from the poles of the unregulated Eisenstein series $E^{SO(d,d)}_{S,s}$ and $E^{SO(d,d)}_{C,s}$ at $s=2$. 

Our first aim in this note will be to give a precise renormalization prescription for the integral \eqref{dfourrfour2}, which is divergent when $d\geq 3$, and show that it indeed satisfies the differential equation \eqref{delsod4r4}, with the correct coefficients of the harmonic anomalies. The renormalization prescription requires a careful treatment of the contributions from degenerate Riemann surfaces, corresponding to primitive two-loop divergences, one-loop subdivergences and overlapping subdivergences. From the proof it will transpire that the anomalous terms on the right-hand side of  \eqref{delsod4r4} originate from these degenerations. 

Unlike the  $\cR^4$ and $D^4\cR^4$ couplings,  the next term in the low-energy expansion of the four-graviton scattering amplitude, namely the $D^6\cR^4$ coupling $\cE^{(d)}_{(0,1)}$, is not a residue of Langlands-Eisenstein series for the U-duality group. Indeed it must satisfy a U-duality invariant Laplace-type equation
with a source term proportional to the square of the $\cR^4$ coupling $\cE^{(d)}_{(0,0)}$ 
\cite{Green:2005ba,Green:2010wi,Green:2010sp,Bossard:2015uga,Wang:2015jna}
(up to harmonic anomalies linear in $\cE^{(d)}_{(0,0)}$ and $\cE^{(d)}_{(1,0)}$ in special dimensions, computed in \cite{Pioline:2015yea} and confirmed in \cite{Bossard:2015oxa}). In particular, the two-loop contribution to $D^6 \cR^4$ is given by the modular integral  \cite{D'Hoker:2013eea,D'Hoker:2014gfa}
\be
\label{dsixrfour2}
\cE^{(d,2)}_{(0,1)}(G,B) = \pi\, \RN \int_{\cF_2} \de\mu_2 \, \Gamma_{d,d,2}(\Omega;G,B)\, \varphi(\Omega)\ ,
\ee
where $\varphi(\Omega)$ is the Kawazumi-Zhang invariant, a real-analytic Siegel modular function
introduced in the mathematics literature in   \cite{Kawazumi,zbMATH05661751}. As before, the integral \eqref{dsixrfour2} is divergent when $d\geq 2$, and requires a renormalization prescription. The Laplace-type equation for $\cE^{(d)}_{(0,1)}$ implies that the renormalized two-loop contribution must satisfy\footnote{The harmonic anomaly for $d=2$, unlike for $d=5$ and $d=6$, turns out to depend on the renormalization scheme. It can be removed by adding to $\cE^{(d,2)}_{(0,1)}$ a suitable multiple   of $\cE^{(d,1)}_{(0,0)}$ and a constant.}
\be
\label{delsod6r4}
\begin{split}
\left( \Delta_{SO(d,d)}  - (d+2)(5-d) \right)\, \cE_{(0,1)} ^{(d,2)}  =&  - \left ( \cE_{(0,0)} ^{(d,1)} \right )^2 -\left(\frac{\pi}{3} \cE_{(0,0)}^{(2,1)} + \frac{7\pi^2}{18} \right) \, \delta_{d,2}  \\
& + \frac{70}{3}\zeta(3) \delta_{d,5} + \frac{20}{\pi}  \cE_{(1,0)}^{(6,1)}  \delta_{d,6} \;,
\end{split}
\ee
where $\cE_{(1,0)} ^{(d,1)}$ are is the one-loop contributions to the
 $D^4\cR^4$ couplings,
 \be
 \label{r4oneloopb}
\cE_{(1,0)} ^{(d,1)}(G,B)  = 2\pi \, \RN \int_{\cF_1} \de \mu_1 \, \Gamma_{d,d,1}(\tau;G,B) \, 
E^\star(2,\tau)\ ,
\ee
where $E^\star(s,\tau)$ is the non-holomorphic Eisenstein series for $SL(2,\IZ)$, normalized as in \cite{D'Hoker:2014gfa}. The
appearance of the quadratic term $-  ( \cE_{(0,0)} ^{(d,1)} )^2$ on the r.h.s. of \eqref{delsod6r4} makes it clear that $ \cE_{(0,1)} ^{(d,2)}$ cannot be a residue of a Langlands-Eisenstein series. Indeed, a candidate for the non-perturbative completion of the $D^6\cR^4$ couplings is only available for 
$d\leq 4$ \cite{Green:2005ba,Basu:2007ck,Green:2010wi,Green:2014yxa,Basu:2014hsa,Pioline:2015yea}.
As for the modular integral \eqref{dfourrfour2}, we shall give a precise renormalization prescription for 
the modular integral \eqref{dsixrfour2}, and establish the differential equation \eqref{delsod6r4} by a careful analysis of the contributions from degenerate Riemann surfaces. In particular, it will transpire that the quadratic term on the r.h.s. of \eqref{delsod6r4} originates from a logarithmic singularity of the Kawazumi-Zhang invariant $\varphi(\Omega)$ in the separating degeneration limit, while the remaining terms originate from primitive two-loop divergences and one-loop subdivergences. 

It is important to stress that these results depend on essential properties of the Kawazumi-Zhang 
invariant, which were originally guessed by trying to derive the differential equation \eqref{delsod6r4}
from the modular integral \eqref{dfourrfour2}, but which have  been since then established independently with mathematical rigor \cite{D'Hoker:2014gfa,Pioline:2015qha}. In particular, the fact that the modular integral \eqref{dfourrfour2} is an eigenmode of the Laplacian $\Delta_{SO(d,d)} $ with eigenvalue $(d+2)(5-d)$, up to harmonic anomalies, strongly pointed to the fact that $\varphi(\Omega)$ had to be an eigenmode of the Laplacian
$\Delta_{Sp(4)}$ on the Siegel upper half plane of degree 2 with eigenvalue 5 \cite{D'Hoker:2014gfa}. Similarly, the fact that logarithmic divergences occur only in $d=2,5,6$ was a strong indication about the asymptotics of $\varphi(\Omega)$ in the non-separating degenerations, eventually leading to the discovery of the
Theta lift representation of the Kawazumi-Zhang invariant  \cite{Pioline:2015qha}.

The outline of this work is has follows. In \S\ref{sec:rencoup}, we give a precise renormalization prescription for the modular integrals \eqref{dfourrfour2}, \eqref{r4oneloopa}, \eqref{dsixrfour2}, \eqref{r4oneloopb}, which are naively divergent for large enough values of the dimension $d$. 
The renormalization of the one-loop amplitudes \eqref{r4oneloopa}, \eqref{r4oneloopb} is standard,
but the renormalization of the two-loop amplitudes \eqref{dfourrfour2} and \eqref{dsixrfour2} 
requires a careful treatment of the minimal and maximal non-separating degenerations.
In \S\ref{sec:laplace}, we establish the differential equations satisfied by these renormalized couplings, and compute the precise coefficients of the harmonic anomalies, confirming the values predicted
by U-duality. In \S\ref{sec:discussion} we close with some open questions. 

\medskip

Note added: while this article was being finalized, we received the preprint \cite{Basu:2015dqa},
which has some overlap with the present work.

\section{Renormalised couplings} \label{sec:rencoup}

The couplings $\cE^{(d)}_{(m,n)} D^{4m+6n}\cR^4$ of interest in this work refer to local terms in the
low-energy expansion of the one-particule irreducible effective action of type II string theory compactified on a torus $T^d$. In dimension $D=10-d>4$, the 1PI effective action is finite, both in the ultraviolet and in the infrared. Due to massless thresholds however, it is a non-analytic function of the momenta. In order to isolate the local part of the effective action, it is convenient to introduce an infrared cut-off $\Lambda$ to separate the contribution of massless supergravity states from those of massive string states, and take the low-energy expansion of each parts separately \cite{Green:1999pv,
Green:2008uj,Tourkine:2013rda}. The supergravity contribution leads to non-local terms in the effective action, supplemented with a set of local counterterms depending on $\Lambda$, which act as a ultraviolet cut-off for the supergravity modes, while the string theory contribution leads to local interactions only, which also depends on $\Lambda$.  The sum of the string theory and supergravity contributions to the coefficients of the local interation $D^{4m+6n}\cR^4$ has a finite limit as the cut-off $\Lambda$ is removed,  and defines the renormalized 
coupling $\cE^{(d)}_{(m,n)}$.

In more detail, the string theory
contribution to the coefficient of the $D^{4m+6n}\cR^4$ term at $h$-loop is given by 
\be
\label{Fmn}
\int_{\cM_h^\Lambda} \de\mu_h\, F_{(m,n)}^{(d,h)} \, \Gamma_{d,d,h}(\Omega;G,B)
\ee
where $F_{(m,n)}$ is a specific function on the moduli space $\cM_h$ of compact Riemann surfaces of genus $h$. The lattice partition function $\Gamma_{d,d,h}$ is 
\be\label{eq:Hf}
\Gamma_{d,d,h}(\Omega;G,B) = (\det\Omega_2)^{d/2} \sum_{m_i^I \, n^{i,I}\in \IZ^{h d}}
e^{-\pi \mathcal{L}^{IJ} \Omega_{2,IJ} + 2\pi\I m_i^I n^{i,J} \Omega_{1,IJ}}
\ee
where $\mathcal{L}^{IJ}$ is a positive-definite quadratic form in the momentum and winding numbers $m_i^I, n^{i,I}$, $i=1\dots d$, $I=1\dots h$, given in terms  of the metric $G_{ij}$ and Kalb-Ramond two-form $B_{ij}$ on the torus $T^d$ via
\be
\mathcal{L}^{IJ}  = (m_i^I + B_{ij} n^{j,I}) G^{ik} (m_k^J + B_{kl} n^{l,J})
+ n^{i,I} G_{ij} n^{j,J}\ .
\ee
In writing \eqref{Fmn} we have reduced the integral over the moduli space of super-Riemann surfaces of genus $h$ to an integral over $\cM_h$. The integrand is independent of the 
choice of projection up to total derivatives, which we assume do not contribute in this highly supersymmetric set-up. The integration domain $\cM_h^\Lambda$ is a subset of $\cM_h$ which removes a neighborhood of the singular locus in $\cM_h$ where the Riemann surface develops a node,
such that $\lim_{\Lambda\to\infty} \cM_h^\Lambda = \cM_h$. In this limit, the integral \eqref{Fmn} generally grows as finite sum of positive powers of the cut-off $\Lambda$, up to logarithms,
\be
\int_{\cM_h^\Lambda} \de\mu_h\, F_{(m,n)}^{(d,h)} \, \Gamma_{d,d,h}(\Omega;G,B) 
\sim e_{(m,n)}^{(d,h)}(\Lambda;G,B) = \sum_{k=1}^{\ell} a_k(G,B)\, \Lambda^{\alpha_k} (\log \Lambda)^{m_k}\ .
\ee
The coefficients $a_k(G,B)$ are controlled by the behavior of $F_{(m,n)}$ near the singular locus.
Near a separating divisor (relevant for $h>1$ only), $\Sigma_h$ degenerates into the product of two Riemann surfaces $\Sigma_{h'}$ and $\Sigma_{h''}$ with $h=h'+h''$, joined by a long tube. Accordingly, $a_k(G,B)$ will be proportional to the product of two modular integrals over $\cM_{h'}$ and $\cM_{h''}$. Near a non-separating divisor, $\Sigma_h$ degenerates into a Riemann surface $\Sigma_{h-1}$ with two punctures joined by a long tube, and $a_k(G,B)$ is proportional to a modular integral over $\cM_{h-1}$.

The supergravity contribution, corresponding to the integral over the complement
of $\cM_h^\Lambda$ inside $\cM_h$, cancels these power-like terms, leaving a finite coefficient for the term $D^{4m+6n}\cR^4$  in the local effective action
\be
\begin{split}
\cE^{(d)}_{(m,n)} = &
 \lim_{\Lambda\to\infty} \left[ \int_{\cM_h^\Lambda} \de\mu_h\, F_{(m,n)} \, \Gamma_{d,d,h}(\Omega;G,B)  - e_{(m,n)}^{(d,h)}(\Lambda;G,B) \right]
\end{split}
\ee
which defines the renormalized integral $\RN \int_{\cM_h} \de\mu_h\, F_{(m,n)}^{(d,h)} \Gamma_{d,d,h}(\Omega;G,B)$. Notice that the supergravity contribution includes loop diagrams with insertions of counterterms cancelling divergences at lower order in string perturbation theory.

In this paper, our main interest is on the two-loop contributions $\cE^{(d,2)}_{(1,0)}$ 
and $\cE^{(d,2)}_{(0,1)}$. As a warm-up however, we briefly discuss the renormalisation of the one-loop contributions to $\cE^{(d)}_{(0,0)}$ and $\cE^{(d)}_{(1,0)}$, as they also enter as 
subdivergences of the two-loop amplitudes mentioned above. We shall briefly comment on 
three-loop contributions to $\cE^{(d)}_{(0,1)}$ in \S\ref{sec:discussion}.

\subsection{One-loop renormalization}

At one-loop, infrared divergences potentially come from the region $\rho_2\to\infty$ in the 
standard fundamental domain $\cF_1=\{\rho\in \cH_1, |\rho|>1, -\tfrac12<\rho_1\leq \tfrac12\}$. As in
\cite{MR656029,Green:1999pv,Angelantonj:2011br}, they can be regulated by truncating the fundamental domain to  ${\cF_1}^\Lambda = {\cF_1} \cap \{ \rho_2 \leq \Lambda\}$. Note that the measure  is normalized to $\de\mu_1(\rho) = 2 \de\rho_1 \de\rho_2 / \rho_2^2$.
Using the following estimates for large $\rho_2$, valid up to exponentially suppressed corrections,
\be
\begin{split}\label{eq:1loap}
\Gamma_{d,d,1}(\rho,G,B) \sim & ~\rho_2^{d/2}\ , \qquad 
E^\star(s;\rho) \sim  \zetastar(2s)\, \rho_2^s +  \zetastar(2s-1) \rho_2^{1-s}\ ,
\end{split} 
\ee
where $\zetastar(s) = \pi^{-s/2} \Gamma(s/2) \zeta(s)$ satisfies $\zetastar(s)=\zetastar(1-s)$, 
it is straightforward to determine the divergent part of the regulated integrals,
\begin{subequations}
\label{r4oneloopR}
\begin{align} \label{r4oneloopRa}
 \int_{\cF_1^\Lambda} d \mu_1\, \Gamma_{d,d,1} \sim &~ \frac{2\Lambda^{\frac{d}{2}-1}}{\frac{d}{2}-1} \Theta(d-2)+ 2\delta_{d,2} \log\Lambda\ ,  \\
\label{r4oneloopRb}
 \int_{\cF_1^\Lambda} d \mu_1 \, \Gamma_{d,d,1}\, E^\star(2,\rho) \sim & ~
 \frac{ 2\zeta^\star(4) \Lambda^{\frac{d}{2}+1}}{\frac{d}{2}+1} + 
 \frac{2\zeta^\star(3) \Lambda^{\frac{d}{2}-2}}{\frac{d}{2}-2} \Theta(d-4)+ 
2\zeta^\star(3)  \delta_{d,4} \log\Lambda
\end{align}
\end{subequations}
where $\Theta(x)=1$ if $x>0$ and zero otherwise. These divergent parts originate from contributions of massless modes, and are cancelled by the supergravity counterterms. Thus, the renormalised couplings in~\eqref{r4oneloopa}, \eqref{r4oneloopb}, are given by 
\begin{subequations}
\label{r4oneloopR2}
\begin{align} \label{r4oneloopRc}
\cE_{(0,0)} ^{(d,1)}  =&~\lim_{\Lambda\to\infty}\left[ \pi\, \int_{\cF_1^\Lambda} \de \mu_1\, \Gamma_{d,d,1} (\rr;\rho) -2\pi \left(\frac{\Lambda^{\frac{d}{2}-1}}{\frac{d}{2}-1} \Theta(d-2)+ \delta_{d,2} \log\Lambda\right)\right]\;,  \\
\label{r4oneloopRd}
\cE_{(1,0)} ^{(d,1)}  =& \lim_{\Lambda\to\infty}\Bigg[2\pi \, \int_{\cF_1^\Lambda} \de \mu_1 \, \Gamma_{d,d,1}(\rr;\rho) \, E^\star(2,\rho) - 4 \pi 
\\ \nonumber &~~~~\times \left( 
\zeta^\star(4) \frac{\Lambda^{\frac{d}{2}+1}}{\frac{d}{2}+1} + 
\zeta^\star(3) \frac{\Lambda^{\frac{d}{2}-2}}{\frac{d}{2}-2} \Theta(d-4)+ 
\zeta^\star(3)  \delta_{d,4} \log\Lambda \right)\Bigg]\;.
\end{align}
\end{subequations}
This renormalization prescription is a special case of the general method developed 
in \cite{MR656029}.

\subsection{Two-loop renormalization, generalities}

At genus 2, the moduli space of Riemann surfaces can be identified with a fundamental domain $\cF_2$ for the action of the modular group $Sp(4,\IZ)$ on the complement of the separating divisor 
$\cD$ in the Siegel upper-half plane $\cH_2$. The latter is parametrized by the period 
matrix $\Omega$, a symmetric complex valued two-by-two matrix whose imaginary part is positive definite. The separating divisor corresponds to the locus $\Omega_{12}=0$, along with all its images under $Sp(4,\IZ)$. We choose the same  fundamental domain $\cF_2$ as in \cite[A.15]{D'Hoker:2014gfa},
\be
\label{defF2}
\begin{split}
(1) & \qquad -\frac12 < \Re(\Omega_{11}), \Re(\Omega_{12}), \Re(\Omega_{22}) \leq \frac12 \\
(2) & \qquad 0< 2 \Im(\Omega_{12}) \leq \Im(\Omega_{11}) \leq \Im(\Omega_{22})     \\
(3) & \qquad  |\det(C\Omega+D)|>1 \ \mbox{for all}\  \begin{pmatrix} A & B \\ C& D \end{pmatrix} \in Sp(4,\IZ)
\end{split}
\ee
Infrared divergences originating from the separating degeneration can be regulated by enforcing a
cut-off $|\Omega_{12}|>\epsilon$. As we shall see below, the modular integrals \eqref{dfourrfour2}
and \eqref{dsixrfour2} are in fact convergent in this region, but the action of the Laplace operator $\Delta_{SO(d,d)}$ on the integrand of \eqref{dsixrfour2} renders the integral divergent, and is responsible for the quadratic anomalous term on the r.h.s. of \eqref{delsod6r4}.

For what concerns the non-separating degeneration limit, it is useful to parametrize the  period matrix as follows:
\be
\label{Omparamrho}
\Omega = \begin{pmatrix}
    \rho & u_1 + \rho u_2  \\ u_1 + \rho u_2  & \sigma_1 + \I(t + \rho_2 u_2^2)
    \end{pmatrix}\ ,
\ee
where $\rho$ is a complex modulus in the Poincar\'e upper half-plane $\cH_1$, $t\in \IR^+$ and $u_1,u_2,\sigma_1$ are real. The non-separating degeneration limit corresponds to $t\to\infty$ keeping the  other variables fixed. In this region, the inequalities \eqref{defF2} defining the 
fundamental domain reduce to 
\be
\label{F2simp}
0<u_2\leq \tfrac12\ ,\quad -\tfrac12<\rho_1,u_1,\sigma_1<\tfrac12\ ,\quad
|\rho|^2>1\ , \quad  \rho_2 (1-u_2^2)\leq t \ ,\quad t>0\ .
\ee
In particular, $\rho$ takes values in the one-loop fundamental domain $\cF_1$ and $\rho_2$ cannot exceed $4t/3$. To regulate potential divergences from the non-separating degeneration, it is therefore sufficient to truncate the integration domain to ${\cF_2}^\Lambda = {\cF_2} \cap \{ t \leq \Lambda\}$.

\begin{figure}
\centerline{\hfill\raisebox{2ex}{\includegraphics[height=6cm]{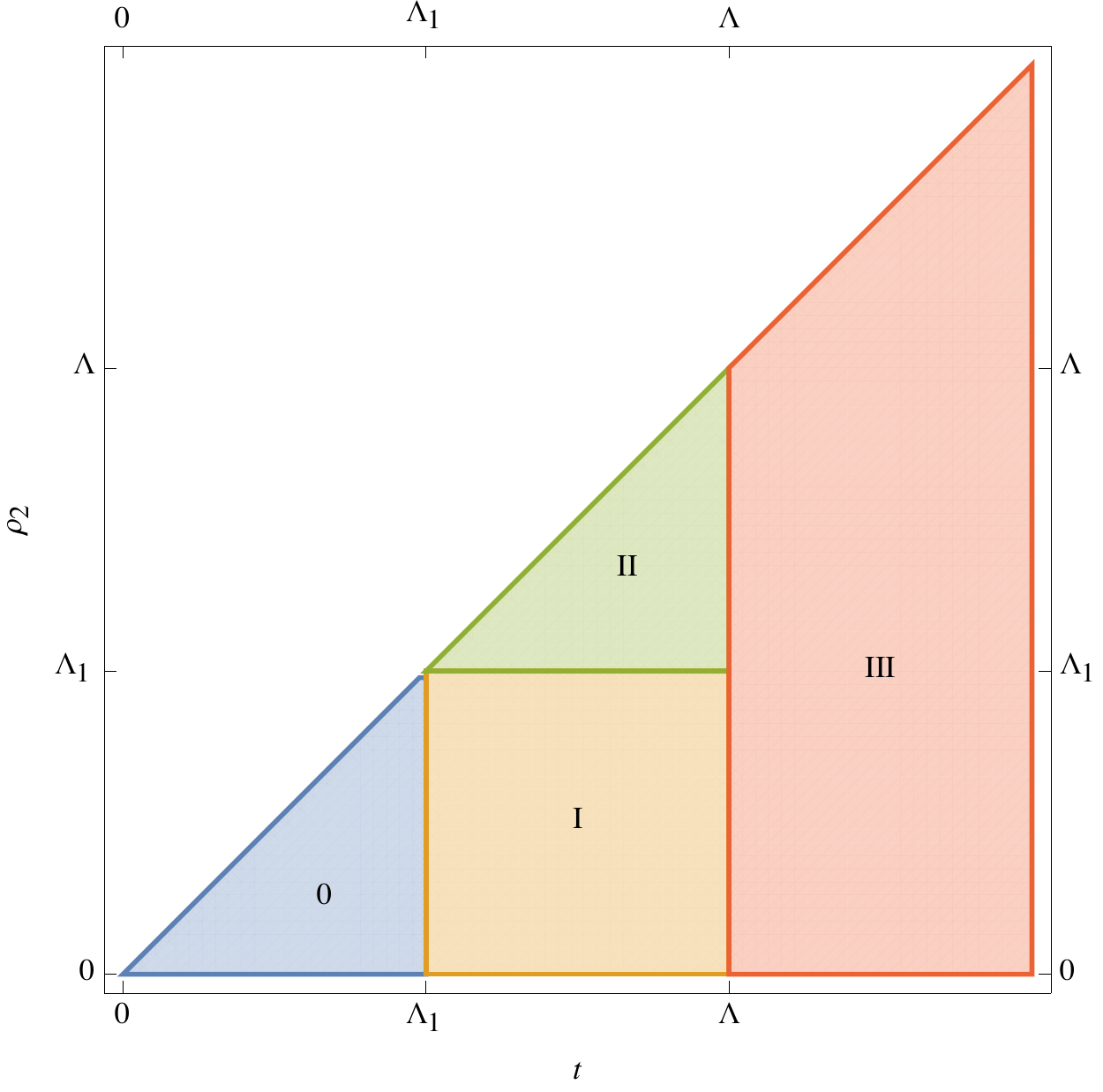}}\hfill
\includegraphics[height=6.6cm]{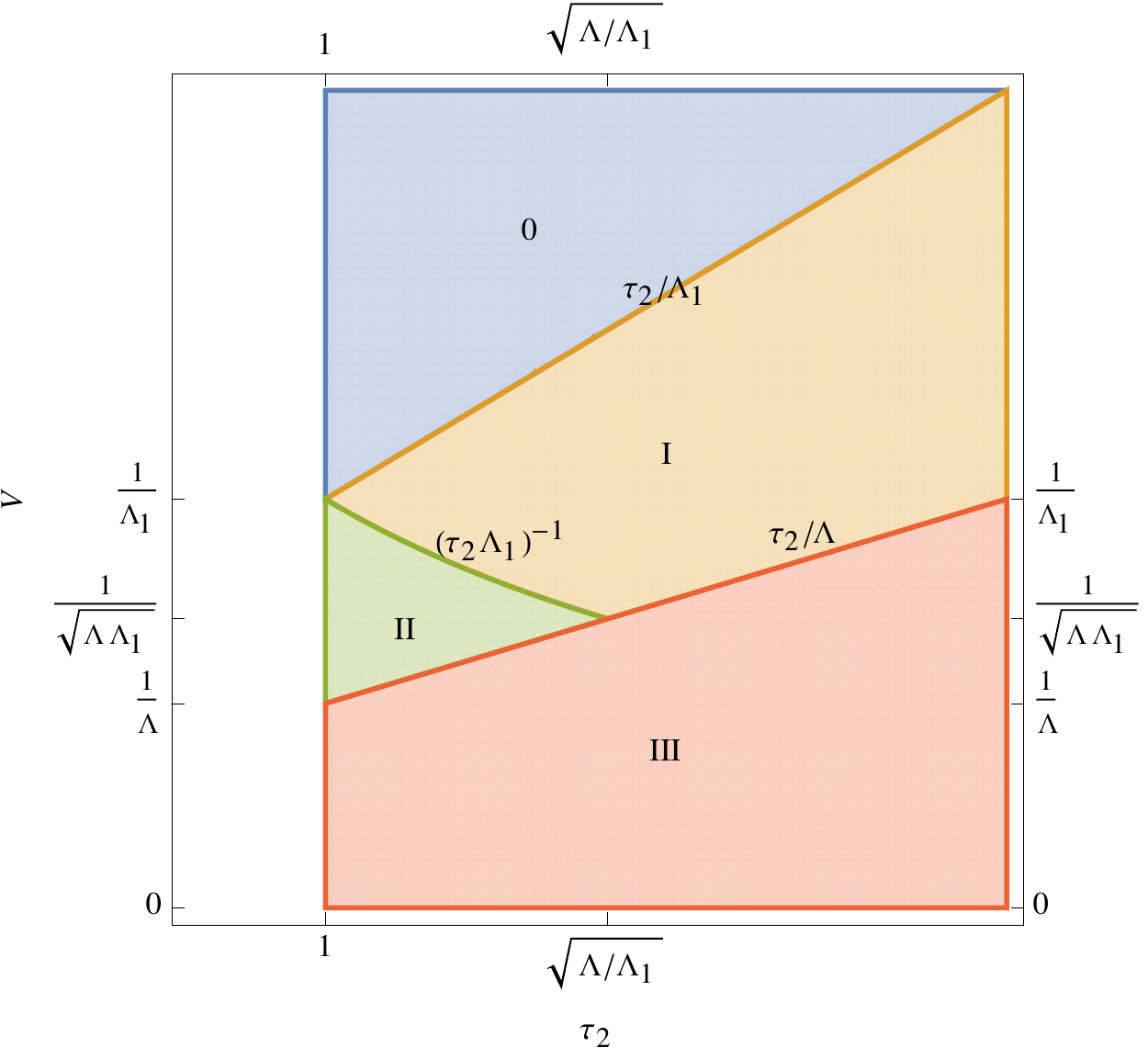}\hfill}
\caption{The cut-off fundamental domain domain $\cF_2^{\Lambda}$ and its  splitting into regions 0, I and II is depicted in $(t,\rho_2)$ coordinates (left) and $(V,\tau_2)$ coordinates, assuming that the off-diagonal entry $\tau_1=\rho_2 u_2$ vanishes. Region III denotes the complement of $\cF_2^{\Lambda}$ inside $\cF_2$.\label{figreg}}
\end{figure}

To disentangle the contributions from the minimal non-separating degeneration limit, where the Riemann surface develops only one non-separating node, from the maximal non-separating degeneration limit or leading singularity, where the Riemann surface develops three non-separating nodes, it  is useful to further split ${\cF_2}^\Lambda$ into three regions (see Figure \ref{figreg}):
\be
\label{defregIII}
\begin{split}
{\cF_2}^{0}= & {\cF_2}^\Lambda \cap \{ \rho_2 \leq  t+u_2^2\rho_2 \leq \Lambda_1 \} \ , \\
{\cF_2}^{I}= & {\cF_2}^\Lambda \cap \{ \rho_2 \leq \Lambda_1 \leq t+u_2^2\rho_2 \} \ , \\
{\cF_2}^{II}= &{\cF_2}^\Lambda \cap \{ \Lambda_1 \leq \rho_2 \leq   t+u_2^2\rho_2\}\ ,
\end{split}
\ee
where $\Lambda_1$ regulates the infrared divergences associated to the coefficient of the one-loop subdivergence (also known as overlapping divergences). The sum of the contributions of the three regions is of course independent of $\Lambda_1$, while mixed terms depending on both $\Lambda$ and $\Lambda_1$ cancel in the sum of regions I and II.

To describe the contributions from the region ${\cF_2}^{II}$, which is associated to primitive two-loop divergences, it is convenient to use yet a different set of variables for the imaginary part of the
period matrix, introduced in \cite{Green:1999pu,Green:2005ba,Green:2008bf}, 
\be
\label{Omparamtau}
\Omega = \begin{pmatrix} \rho_1 & u_1 \\ u_1 & \sigma_1 \end{pmatrix} 
+ \frac{\I}{\tau_2 V}\begin{pmatrix}  1 & \tau_1 \\ \tau_1 & |\tau|^2 \end{pmatrix}
\ee
The two parametrizations \eqref{Omparamrho} and \eqref{Omparamtau} are related by 
\be
V=\frac{1}{\sqrt{t \rho_2}}\ ,\quad \tau_2 = \sqrt{\frac{t}{\rho_2}}\ ,\quad \tau_1 = u_2\ ,
\ee
while the integration measure in either set of variables reads
\be
\de\mu_2(\Omega) = 4 \frac{\de t}{t^3}\, \de \mu_1(\rho)\, \de u_1\, \de u_2 \, \de \sigma_1 =
8 V^2 \de V\, \de \mu_1(\tau) \, \de \rho_1\, \de u_1\, \de \sigma_1\ .
\ee
In the region $V\to 0$, where all entries in $\Omega_2$ are scaled to infinity at the same rate, the inequalities \eqref{defF2} defining the fundamental domain $\cF_2$ reduce to 
\be\label{fd2l}
0<\tau_1 \leq \tfrac12\ ,\quad |\tau|^2\geq 1\ ,\quad -\tfrac12 < \rho_1, u_1, 
\sigma_1 \leq \tfrac12\ ,\quad V>0\ ,
\ee
so that $\tau$ lies in the fundamental domain $\cF_1/\IZ_2$ of the action of $GL(2,\IZ)$ on $\cH_1$ (the latter consisting of the usual fractional linear transformations of $\tau$, along with the involution $\IZ_2:\tau\to-\bar\tau$). The region II of the truncated fundamental domain $\cF_2^\Lambda$ enforces two additional
inequalities,
\be
\cF_2^{II}= \cF_2 \cap \left\{   \tau_2 \leq \sqrt{\frac{\Lambda}{\Lambda_1}}\ ,\quad 
 \frac{\tau_2}{\Lambda} < V < \frac{1}{\tau_2 \Lambda_1} \right\}\ .
\ee
In particular, $V$ is bounded from below by $\sqrt{3}/(2\Lambda)$ and from above by $\sqrt{2}/(3\Lambda_1)$.

For later reference, we compute, for $\alpha\neq -3$, $\alpha+\beta\neq-2$ and $\alpha-\beta\neq -4$, under the assumption that $\Lambda_1$ is large enough so that the inequalities defining the fundamental domain $\cF_2$ simplify to \eqref{F2simp},
\be
\label{intLI}
\begin{split}
\int\limits_{\cF_2^{0}\cup\cF_{2}^{I}} \!\!\! \de\mu_2 V^{\alpha} \tau_2^\beta = & 
~8 \int_0^{1/2} \de u_2 \, 
\int_{-1/2}^{1/2} \de\rho_1
\int_{\sqrt{1-\rho_1^2}}^{\Lambda_1} \rho_2^{-\frac{\alpha+\beta}{2}-2} \de\rho_2
\int_{\rho_2 (1-u_2^2)}^\Lambda t^{\frac{\beta-\alpha}{2}-3} \de t
\\
 = &  \frac{16\Lambda_1^{-\frac{\alpha+\beta+2}{2}}\Lambda^{\frac{\beta-\alpha-4}{2}}}{(\alpha+\beta+2)(4+\alpha-\beta)}
-\frac{16 c(\frac{\alpha-\beta+4}{2}) \Lambda_1^{-3-\alpha}}{(\alpha+3)(4+\alpha-\beta)}  \\
&
-\frac{32 c(\frac{\alpha+\beta+2}{4}) \Lambda^{\frac{\beta-\alpha-4}{2}}}{(4+\alpha-\beta)(\alpha+\beta+2)}
+\frac{32c(\frac{\alpha-\beta+4}{4})\, c(\frac{\alpha+3}{2})}{(\alpha+3)(4+\alpha-\beta)}
\end{split}
\ee
\be
\label{intLII}
\begin{split}
\int_{\cF_2^{II}} \de\mu_2 V^{\alpha} \tau_2^\beta = & 
~ 16 \, \int_0^{1/2}\de\tau_1\,
\int_{\sqrt{1-\tau_1^2}}^{\sqrt{\Lambda/\Lambda_1}}  \tau_2^{\beta-2}\, \de\tau_2\,
\int_{\tau_2/\Lambda}^{1/(\tau_2\Lambda_1)}  V^{2+\alpha} \de V\,  
\\
=& \frac{16c(-\frac{\alpha+\beta+2}{2}) \Lambda^{-3-\alpha}}{(\alpha+3)(\alpha+\beta+2)}  
+ \frac{16c(\frac{\alpha-\beta+4}{2}) \Lambda_1^{-3-\alpha}}{(\alpha+3)(4+\alpha-\beta)}  
- \frac{16\Lambda_1^{-\frac{\alpha+\beta+2}{2}} \Lambda^{\frac{\beta-\alpha-4}{2}}}{
(\alpha+\beta+2)(4+\alpha-\beta)}
\end{split}
\ee
where 
\be
c(\gamma) = \int_0^{\frac{1}{2}} (1-x^2)^{-\gamma} \de x = \frac{1}{2}\, 
{}_2 F_1 \left(\gamma,\tfrac{1}{2};\tfrac{3}{2};\tfrac{1}{4}\right)\ .
\ee
We note the special values $c(0)=\frac12, c(\tfrac12)=\frac{\pi}{6}, c(-1)=\frac{11}{24}$.
As expected, the $\Lambda_1$ dependence cancels in the sum, leaving only
\be
\label{intLtot}
\int_{\cF_2^\Lambda} \de\mu_2 V^{\alpha} \tau_2^\beta \sim 
-\frac{32 c(\frac{\alpha+\beta+2}{4}) \Lambda^{\frac{\beta-\alpha-4}{2}}}{(4+\alpha-\beta)(\alpha+\beta+2)}
+\frac{16c(-\frac{\alpha+\beta+2}{2}) \Lambda^{-3-\alpha}}{(\alpha+3)(\alpha+\beta+2)} \ ,
\ee 
where we have neglected $\Lambda$-independent terms.
It is worth noting that the $\cO(\Lambda^{\frac{\beta-\alpha-4}{2}})$ term in \eqref{intLtot}
originates from the boundary at $t=\Lambda$ in \eqref{intLI}, while the $\cO( \Lambda^{-3-\alpha})$
originates from the boundary at $V=\tau_2/\Lambda$ in \eqref{intLII}.

\subsection{Renormalized \texorpdfstring{$D^4\cR^4$}~~coupling~at two-loop}
\label{sect:rend4r4}

We are now ready to compute the divergent part of the modular integral \eqref{dfourrfour2}.  In region I where $t\gg \rho_2$, it is clear from \eqref{eq:Hf} that  the lattice partition function can be 
approximated as
\be
\label{approxI}
\Gamma_{d,d,2} \sim t^{d/2} \Gamma_{d,d,1}(\rho)\ ,
\ee
up to exponentially suppressed corrections in $\Lambda$. Thus we have
\be
\frac{\pi}{2} \int_{{\cF_2}^{I}} \!\!\!\de\mu_2\, \Gamma_{d,d,2} \sim 
2\pi \int_0^{1/2} \!\! d u_2 \, 
\int_{\cF_1^{\Lambda_1}} \!\!\!\de\mu_1(\rho)\, \Gamma_{d,d,1}(\rho)\,
\int_{\rho_2 (1-u_2^2)}^\Lambda t^{\frac{d}{2}-3}  \de t\ .
\ee
Using \eqref{r4oneloopRc} and focusing only on the divergent contributions as $\Lambda \to \infty$ we have
\be
\label{d4r4reg1}
\frac{\pi}{2} \int_{{\cF_2}^{I}} \!\!\!\de\mu_2\, \Gamma_{d,d,2} \sim  
\left[ \frac{ \Lambda^{\frac{d}{2}-2}}{\tfrac{d}{2}-2} \Theta(d-4) + \log \Lambda\, \delta_{d,4}\right]
\left [ \cE_{(0,0)}^{(d,1)} 
+ \frac{4 \Lambda_1^{\tfrac{d}{2}-1}}{d-2} \Theta(d-2) \right]
\ee

In region II, where all entries of $\Omega_2$ are large, we can instead approximate 
\be
\label{approxII}
\Gamma_{d,d,2} \sim (\det\Omega_2)^{d/2} = V^{-d}\ ,
\ee 
corresponding to the contributions of the massless supergravity modes. Using \eqref{intLII}, we find
\be
\label{d4r4reg2}
\begin{split}
\frac{\pi}{2} \int_{{\cF_2}^{II}} \de\mu_2\, \Gamma_{d,d,2} \sim  &
 \frac{8\pi \, c(\frac{d}{2}-1) \Lambda^{d-3}}{(d-2)(d-3)}\, \Theta(d-3)+ \frac{4\pi^2}{3} \log\Lambda \, \delta_{d,3}  \\ &  
 -\frac{8\pi \Lambda^{\frac{d-4}{2}}  \Lambda_1^{\frac{d-2}{2}}}{(d-2)(d-4)}\, \Theta(d-4) - 2\pi \delta_{d,4}\, \Lambda_1 \log\Lambda
\end{split}
\ee
As expected, the terms depending on both $\Lambda_1$ and $\Lambda$, corresponding to overlapping divergences, cancel in the sum of the contributions of regions I and II. The $\Lambda$-dependent terms, on the other hand, must cancel against the counterterms. The term proportional to $\Lambda^{\tfrac{d}{2}-2}$ in \eqref{d4r4reg1} corresponds to a one-loop subdivergence, while the term proportional to $\Lambda^{d-3}$, which originates from the boundary $V=\tau_2/\Lambda$ in the integral over $V$
in \eqref{d4r4reg2}, corresponds to the primitive two-loop divergence. Its coefficient is recognized as $4\pi \cI_d/(d-3)$, where $\cI_d$ is the renormalized integral
\be
\label{defId}
\cI_d= \RN \int_{\cF_1/\IZ_2} \, \de\mu(\tau)\, \tau_2^{3-d} = \frac{2\, c(\tfrac{d}{2}-1)}{d-2} \ .
\ee
This integral converges for $d>2$, and its renormalized value is defined for any $d\neq 2$ by analytic continuation\footnote{In \eqref{defId},  $\tau_2^{3-d}$ denotes the modular invariant (but not smooth) function which is equal to $\tau_2^{3-d}$ in the fundamental domain $\cF_1/\IZ_2$.}.
The renormalized $D^4 \cR^4$ coupling at two-loop is defined by subtracting these divergent terms,
\begin{equation}
\label{dfourrfour2R}
\begin{split}
\cE^{(d,2)}_{(1,0)} =& \lim_{\Lambda\to\infty}\left[ \frac{\pi}{2}\, \int_{\cF_2^\Lambda} \de\mu_2 \, \Gamma_{d,d,2}(\Omega) -e^{(d,2)}_{(1,0)} \right] \ ,\\
e^{(d,2)}_{(1,0)}  = &
 \frac{\Lambda^{\frac{d}{2}-2}}{\frac{d}{2}-2} \cE_{(0,0)} ^{(d,1)} \Theta(d-4) 
+ \log\Lambda \, \cE_{(0,0)} ^{(4,1)} \delta_{d,4}  
  + \frac{4 \pi\,  \cI_d\, \Lambda^{d-3}}{d-3}\, \Theta(d-3)
  + \frac{4\pi^2}{3} \delta_{d,3} \log\Lambda  \ .
\end{split}
\end{equation}

\subsection{Renormalized \texorpdfstring{$D^6\cR^4$}~~coupling at two-loop}

In order to compute the divergent part of the two-loop $D^6\cR^4$ coupling \eqref{dsixrfour2}, we need to control the behavior of the Kawazumi-Zhang invariant $\varphi(\Omega)$ in the various degeneration limits. In the separating degeneration $v = u_1 + \rho u_2 \to 0$, one has~\cite{MR1105425,zbMATH06355718}
\be
\label{kzsepdeg}
\varphi(\Omega) = -\log \left| 2\pi v \, \eta^2(\rho)  \eta^2(\sigma) \right| +\cO(|v|^2 \log |v|)\ .
\ee
Fortunately, this logarithmic singularity is integrable, so for the purpose of defining the renormalized integral \eqref{dsixrfour2}, we do not need to excise the region near $v=0$ (however this will be necessary for establishing the differential equation \eqref{delsod6r4}).

The complete asymptotic expansion of $\varphi(\Omega)$ in the non-separating degeneration was established in \cite{Pioline:2015qha}, based on a representation of $\varphi(\Omega)$ as a one-loop modular integral of an almost weakly holomorphic Jacobi form times a lattice partition function of signature (3,2). The upshot of this analysis is that, in the minimal non-separating degeneration $t\to\infty$, the Kawazumi-Zhang invariant behaves as 
\be
\label{kzmindeg2}
 \varphi(\Omega) =  \frac{\pi}{6} t + \varphi_0 + \frac{\varphi_1}{t} + \cO(e^{-t})\ ,
\ee
where
\be
\label{phi1D22}
\varphi_0= \frac12\, \cD_{1,1}(\rho;u_1,u_2)\ ,\quad 
\varphi_1 = \frac{5}{16\pi^2\rho_2}\, \cD_{2,2}(\rho;u_1,u_2) +\frac{5}{2\pi}\, E^{\star}(2;\rho)
\ee
are expressed in terms of the non-holomorphic Eisenstein series $E^{\star}(s;\rho)$ and the Kronecker-Eisenstein series 
\begin{equation}
\label{KronEisZ}
\cD_{a,b} (\rho ; u_1,u_2 ) \equiv \frac{(2\I \rho_2)^{a+b-1}}{2\pi\I}\, \sum_{(m,n)\neq(0,0)}\, \frac{e^{2\pi\I (n\, u_2 + m \, u_1 )}}{(m\rho + n)^a (m\bar\rho+n)^b}\ .
\end{equation}
Importantly, the integrals of $\cD_{a,b} (\rho ; u_1 , u_2 )$ with respect to $u_1$ and $u_2$ in the domain~\eqref{fd2l} vanish when $a+b$ is even. Using the approximation \eqref{approxI} for 
$\Gamma_{d,d,2}$, the divergent part of the integral over region I is therefore
\begin{align}
\nonumber
\pi \int_{{\cF_2}^{I}} \de\mu_2\, \Gamma_{d,d,2}\, \varphi \sim & 
~2 \pi  \int_{\cF_1^{\Lambda_1}} \de\mu_1(\rho) 
\int^\Lambda \! \de t \, t^{\frac{d}{2}-3} \,
\Gamma_{d,d,1}(\rho)\, \left( \frac{\pi t}{6} + \frac{5}{2\pi t} E^{\star}(2;\rho) \right) 
\\ \label{d6r4r1}
\sim & ~ \frac{\pi}{3} \frac{\Lambda^{\frac{d}{2}-1}}{\frac{d}{2}-1}\, 
\left(
\cE_{(0,0)} ^{(d,1)} 
+ 2 \pi \frac{\Lambda_1^{\frac{d}{2}-1}}{\frac{d}{2}-1} \,
\right)
\\ \nonumber
+ & ~ \frac{5}{\pi} \frac{\Lambda^{\frac{d}{2}-3}}{\frac{d}{2}-3}
\left(
\frac{1}{2} \cE_{(1,0)} ^{(d,1)} 
+ 4\pi \zetastar(4)\frac{\Lambda_1^{\frac{d}{2}+1}}{d+2} \,
+ 2 \pi \zetastar(3)\frac{\Lambda_1^{\frac{d}{2}-2}}{\frac{d}{2}-2} \,
\right)\,.
\end{align}
Since we focused on the divergent terms as $\Lambda\to\infty$, one should read this equation disregarding the values of $d$ that yield a negative power of $\Lambda$; the values of $d^*$ for which we have $\Lambda^{\alpha(d^*)}/\alpha(d^*)$ with a vanishing  denominator should be interpreted as limits $d \to d^*$ where only the finite terms are kept. This produces terms that depends on the logarithm of the cutoffs. We will reinstate explicitly the conditions on $d$ and the logarithmic terms in the final result.

In region II, one has instead~\cite{Pioline:2015qha}
\be\label{KZregII}
 \varphi(\Omega) = \frac{\pi}{6V} A(\tau)
 + 
 \frac{5\zeta(3)V^2}{4\pi^2} + \cO(e^{-1/V})
\ee
where
\be
\label{defA}
A(\tau) = \frac{|\tau|^2-\tau_1+1}{\tau_2}+ 5 \frac{(\tau_1^2-\tau_1)(|\tau|^2-\tau_1)}{\tau_2^3} \ .
\ee
Using the approximation \eqref{approxII} for $\Gamma_{d,d,2}$, each term in the integrand reduces to the following generalization of~\eqref{intLII},
\begin{align}
\int_{\cF_2^{II}} \de\mu_2 V^{\alpha}\, \tau_2^\beta \, \tau_1^{2n} = & ~ 
16 \, \int_0^{1/2}\de\tau_1\, \tau_1^{2n} \, 
\int_{\sqrt{1-\tau_1^2}}^{\sqrt{\Lambda/\Lambda_1}}  \tau_2^{\beta-2}\, \de\tau_2\,
\int_{\tau_2/\Lambda}^{1/(\tau_2\Lambda_1)}  V^{2+\alpha} \de V\,  
\nonumber \\
 = & ~ \frac{16c_n(-\frac{\alpha+\beta+2}{2}) \Lambda^{-3-\alpha}}{(\alpha+3)(\alpha+\beta+2)}  
+ \frac{16c_n(\frac{\alpha-\beta+4}{2}) \Lambda_1^{-3-\alpha}}{(\alpha+3)(4+\alpha-\beta)} 
\\ \nonumber
 & - \frac{2^{-2n}}{2n+1} \frac{16\Lambda_1^{-\frac{\alpha+\beta+2}{2}} \Lambda^{\frac{\beta-\alpha-4}{2}}}{(\alpha+\beta+2)(4+\alpha-\beta)}~.
\end{align}
where 
\begin{equation}
  \label{eq:gcg}
   c_n(\gamma) = \int_0^{\frac{1}{2}} x^{2 n} (1-x^2)^{-\gamma} \de x = \frac{4^{-n-1}}{n+\frac{1}{2}}\, {}_2 F_1 \left(\gamma,n+\tfrac{1}{2};n+\tfrac{3}{2};\tfrac{1}{4}\right)\ .
\end{equation}
As clear from~\eqref{KZregII}, we are interested in the cases where either $n$ or $n+\tfrac12$ is integer. In this second case the hypergeometric function in~\eqref{eq:gcg} reduces to an elementary function, while in the first case we can use  
\begin{equation}
  \label{eq:contigrel}
  {}_2 F_1 \left(a,c;c+1;z\right) = \frac{c}{z(c-a)}~ {}_2 F_1 \left(a,c-1;c;z\right) - \frac{c}{z(c-a)}(1-z)^{1-a}
\end{equation}
to collect the different contribution in terms of $c_0(\gamma)\equiv c(\gamma)$. Thus we have
\begin{align}
\pi \!\!\int\limits_{{\cF_2}^{II}} \!\! \de \mu_2\, \Gamma_{d,d,2}\, \varphi ~ \sim ~  & 
 -\frac{2 \pi^2 \Lambda_1^{\frac{d}{2}+1}}{9 (d+2)} 
\frac{\Lambda^{\frac{d}{2}-3}}{\frac{d}{2}-3}
-\frac{8 \pi^2}{3} \frac{\Lambda^{\frac{d}{2}-1}}{(d-2)} 
\frac{\Lambda_1^{\frac{d}{2}-1}}{(d-2)} 
+\frac{4\pi^2 \cI'_d\, \Lambda^{d-2}}{3(d-2)}
 \nonumber \\ \label{d6r4r2}  & 
+ \frac{20\zeta(3)\, c(\tfrac{d}{2}-2)\, \Lambda^{d-5}}{\pi(d-4)(d-5)} - 
5 \zeta(3) \frac{\Lambda^{\frac{d}{2}-3}}{\frac{d}{2}-3}  \frac{\Lambda_1^{\frac{d}{2}-2}}{\frac{d}{2}-2}
\end{align}
where the coefficient of $\Lambda^{d-2}$ is proportional to
\be
\label{defIdp0}
\cI'_d =  \frac{4 (d-2)  c\left(\frac{d}{2}+1\right)}{(d-1) (d+2)} + 
  \frac{ 6(3 d-4) \left(-\frac{d-2}{3}(\frac{4}{3})^{\frac{d}{2}} + \frac{2}{3} (d-1)\right)}{(d-2) (d-1) d (d+2)}\ .
\ee
As in \eqref{defId}, this term originates from the boundary at $V=\tau_2/\Lambda$ in the integral over $V$. The coefficient $\cI'_d$ is recognized as the renormalized integral
\be
\begin{split}
\label{defIdp}
\cI'_d = &\RN \int_{\cF_1/\IZ_2} \de\mu_1(\tau)\, \tau_2^{2-d}\, A(\tau)\ .
\end{split}
\ee
Here the integral converges absolutely for $d>2$, and its renormalized value for $d<2, d\neq -2$ is defined by analytic continuation in $d$. Note that $\cI'_d$ has simple poles at $d=2$, $d=-2$, but  is finite at $d=0$ and $d=1$, since the apparent poles in \eqref{defIdp} cancel.  For future reference, we record the behavior around $d=2$,  $\cI'_d=\tfrac{1}{d-2}+\tfrac{1}{12}+\cO(d-2)$.

It is straightforward to check that the divergent terms depending on both $\Lambda$ and $\Lambda_1$, corresponding to overlapping divergences, cancel after summing~\eqref{d6r4r1} and~\eqref{d6r4r2}. As mentioned after~\eqref{d6r4r1}, the power-like divergences become logarithmic divergences for the values of $d$ where the coefficient has a pole. The renormalized $D^6\cR^4$ two-loop coupling is then defined by subtracting
the divergent terms, 
\be
 \label{dsixrfour2R}
\cE^{(d,2)}_{(0,1)} = \lim_{\Lambda\to\infty} \left[
\pi\, \int_{\cF_2^\Lambda} \, \de\mu_2 \varphi(\Omega)\, \Gamma_{d,d,2}(\Omega)
- e^{(d,2)}_{(0,1)} \right]
\ee
where
\begin{align}
e^{(d,2)}_{(0,1)}= &
\frac{\pi}{3} \frac{\Lambda^{\tfrac{d}{2}-1}}{\frac{d}{2}-1}  \cE_{(0,0)} ^{(d,1)}\, \Theta(d-2) 
+\frac{\pi}{3} \log\Lambda \, \delta_{d,2} \,\cE_{(0,0)} ^{(2,1)} 
\nonumber \\
& + \frac{5}{2\pi}  \frac{\Lambda^{\tfrac{d}{2}-3}}{\frac{d}{2}-3}  \cE_{(1,0)} ^{(d,1)} \, \Theta(d-6) 
+ \frac{5}{2\pi}  \log\Lambda  \, \delta_{d,6}\,  \cE_{(1,0)} ^{(6,1)}
\nonumber \\
  &
+ \frac{10\zeta(3)\,\cI_{d-2}}{\pi(d-5)} \Lambda^{d-5} \, \Theta(d-5)
+\frac{10\zeta(3)}{3} \log\Lambda \, \delta_{d,5} 
\\ \nonumber
&
+\frac{4\pi^2 \cI'_d\, \Lambda^{d-2}}{3(d-2)} \Theta(d-2)
+ \left( \frac{2\pi^2}{3} (\log\Lambda)^2 + \frac{\pi^2}{9}\log\Lambda\right)\, \delta_{d,2}
\ .
\end{align}

For later use, it will be useful to rewrite the renormalized integral $\cI'_d$ defined in \eqref{defIdp0},\eqref{defIdp},  as follows. Using the fact that the function $A(\tau)$ defined in \eqref{defA} and
the factor $\tau_2^{2-d}$
satisfy
\be
\Delta_\tau A = 12 A\ ,\quad \Delta_\tau \tau_2^{2-d}=(d-1)(d-2)\tau_2^{2-d}\ ,
\ee
where $\Delta_{\tau}=\tau_2^2(\partial_{\tau_1}^2+ \partial_{\tau_2}^2)$ and the first identity holds 
away from the separating boundary at  $\tau_1=0$ \cite{Green:2005ba}, we have
\be
\begin{split}
 \cI'_d= &  \frac{1}{12} \int_{\cF_1/\IZ_2} \de\mu_1(\tau)\, \tau_2^{2-d}\, \Delta_\tau\, A(\tau) \\
 = & \frac{(d-1)(d-2)}{12} \int_{\cF_1/\IZ_2} \de\mu_1(\tau)\,  \tau_2^{2-d}\,  A(\tau) 
 +\frac{1}{12} \int_{\partial (\cF_1/\IZ_2)} \tau_2^{2-d} \star \de A- A \star \de  \tau_2^{2-d}\ .
\end{split}
\ee
The normal derivative $\star \de A$ vanishes on the boundaries at $|\tau|=1$ and $\tau_1=\tfrac12$, while it equals $6\de\tau_2/\tau_2$ on the boundary $\tau_1=0$. The normal derivative $\star\de \tau_2^{2-d}=(d-2) \tau_2^{1-d} \de\tau_1$ vanishes on the boundaries $\tau_1=0$ and $\tau_1=1/2$. Thus we get, from the boundaries $\tau_1=0$ and $|\tau|=1$,
\be
\label{Idbound}
\left(1-\tfrac{(d-1)(d-2)}{12}\right) \cI'_d= \int_1^{\infty} \de \tau_2 \, \tau_2^{1-d} - \frac{(d-2)}{6}
\int_0^{1/2} \de\tau_1\, (1-\tau_1^2)^{\frac{1-d}{2}} A\left(\tau_1,\sqrt{1-\tau_1^2}\right)\, 
\ee 
or equivalently,
\be
\label{Idbound2}
\cI'_d = -\frac{12}{(d-2)(d+2)(d-5)} + \frac{2(d-2)}{(d+2)(d-5)}\, \int_0^{1/2} \de\tau_1\, 
(1-\tau_1^2)^{\frac{1-d}{2}} A\left(\tau_1,\sqrt{1-\tau_1^2}\right)\ .
\ee
The first term in this expression is responsible for the pole of $\cI'_d$ at $d=2$, while the apparent
pole at $d=5$ cancels between the two terms in \eqref{Idbound2}. For general  $d$ the integral 
over $\tau_1$ can be performed by using~\eqref{eq:gcg}.  Rewriting the hypergeometric function (obtained from the terms where $n$ in~\eqref{eq:gcg} is integer) in terms of $c(\tfrac{d}{2}+1)$ by using~\eqref{eq:contigrel} and 
\begin{equation}
  \label{eq:contigrel2}
{}_2 F_1 \left(\tfrac{1}{2},\tfrac{d}{2},\tfrac{3}{2},\tfrac{1}{4}\right) = \frac{d}{d-1}\; 
{}_2 F_1 \left(\tfrac{1}{2},\tfrac{d}{2}+1, \tfrac{3}{2},\tfrac{1}{4} \right) - \frac{1}{d-1}
 \left(\tfrac{4}{3}\right)^{\frac{d}{2}}\; ,
\end{equation}
one recovers \eqref{defIdp}. The decomposition \eqref{Idbound2} will however play an important role when computing the action of the Laplacian in \S\ref{sec_laplaced6r4}.

\section{Laplace equations \label{sec:laplace}}

Having defined the renormalized couplings in any dimension, we now proceed to the derivation of
the differential equations \eqref{delsod4r4} and \eqref{delsod6r4}.  Our strategy is simple: we use the following property of the lattice partition function  \cite{Obers:1999um}
\be
\label{DeltaGamma}
\left(\Delta_{SO(d,d)} - 2 \Delta_{Sp(2h)} + \half d h(d-h-1) \right )\, 
\Gamma_{d,d,h} (\Omega) = 0
\ee
in order to convert the action of the Laplacian $\Delta_{SO(d,d)}$ on $\Gamma_{d,d,h}$ into an action of the Laplacian $\Delta_{Sp(2h)}$. Upon  integration by parts, one recovers a multiple of the original regularized integral, except for boundary contributions from degenerate Riemann surfaces, which are responsible for the anomalous terms on the r.h.s. of \eqref{delsod4r4}  and  \eqref{delsod6r4}.

\subsection{One-loop \texorpdfstring{$\cR^4$}~ and \texorpdfstring{$D^4 \cR^4$}~}

As a warm-up, let us apply this procedure to derive the differential equations satisfied by 
the  renormalized one-loop couplings \cite{Pioline:2015yea}
\begin{subequations}
  \label{delsol1}
  \begin{align}
  \label{delsor4l1}    
  \left( \Delta_{SO(d,d)}  + \frac12 d(d-2) \right)\, \cE_{(0,0)} ^{(d,1)}   =& ~  4\pi\, \delta_{d,2}\ , \\  \label{delsor4l2}    
  \left( \Delta_{SO(d,d)}  + \frac12 (d+2)(d-4) \right)\, \cE_{(1,0)} ^{(d,1)} = &~  12\zeta(3)\, \delta_{d,4}\ .
  \end{align}
\end{subequations}
We focus on the coupling $\cE_{(1,0)} ^{(d,1)}$, whose integrand is slightly more complicated, since the calculation for $\cE_{(0,0)} ^{(d,1)}$ easily follows along the same lines. By using~\eqref{DeltaGamma} in~\eqref{delsor4l2} we obtain
\begin{align}
\nonumber
\left(\Delta_{SO(d,d)}  + \frac12(d+2)(d-4) \right) & \cE_{(1,0)} ^{(d,1)} = 4\pi
\lim_{\Lambda\to\infty}\Bigg[\int\limits_{\cF_1^\Lambda} \de\mu_1  E^\star(2;\rho)  \left( \Delta_{Sp(2)} -2 \right) \Gamma_{d,d,1}(\rho)
\\ & \hspace{-2cm} - \left( (d-4) \zeta^\star(4) \Lambda^{\frac{d}{2}+1} + (d+2) \zeta^\star(3) \Lambda^{\frac{d}{2}-2} \Theta(d-4) \right)\Bigg]~.
\label{23b1}
\end{align}
Upon integrating by parts the action of the Laplacian $\Delta_{Sp(2)}=\rho_2^2(\partial_{\rho_1}^2+\partial_{\rho_2}^2)$ and using $\left( \Delta_{Sp(2)} -s(s-1) \right)E^\star(s;\rho)=0$,  we see that the contribution of the integral on the right-hand side localises on the boundary at $\rho_2=\Lambda$. Recalling that $\de\mu_1=2 \de\rho_1 \de\rho_2/\rho_2^2$, the r.h.s. of \eqref{23b1} can be rewritten as 
\begin{equation}
  \label{23b2}
  \begin{split}
  4 \pi \lim_{\Lambda\to\infty}\Bigg[& 2\int_{-\frac{1}{2}}^{\frac{1}{2}} \de\rho_1\, \Big[E^\star(2;\rho) \partial_{\rho_2}  \Gamma_{d,d,1}(\rho) - \Gamma_{d,d,1}(\rho)  \partial_{\rho_2} E^\star(2;\rho) \Big]_{\rho_2=\Lambda} \\ -& \left( (d-4) \zeta^\star(4) \Lambda^{\frac{d}{2}+1} + (d+2) \zeta^\star(3) \Lambda^{\frac{d}{2}-2} \Theta(d-4) \right)\Bigg] = 12 \zeta(3)\delta_{d,4}\ ,
  \end{split}
\end{equation}
establishing the differential equation~\eqref{delsor4l2}. 

\subsection{Two-loop \texorpdfstring{$D^4 \cR^4$}~\label{sec_laplaced4r4}}

We now turn to the analysis of the differential equation~\eqref{delsod4r4} for the two-loop coupling~\eqref{dfourrfour2R}. Again by using~\eqref{DeltaGamma} we find
\begin{align}
\Big( \Delta_{SO(d,d)} & + d(d-3) \Big) \, \cE_{(1,0)} ^{(d,2)} = \lim_{\Lambda\to\infty} \Bigg[ \pi \int_{\cF_2^\Lambda} \!\de\mu_2\, \Delta_{Sp(4)} \Gamma_{d,d,2}(\Omega)
\nonumber \\ \label{121a} &
- 4\pi d \, \cI_d\, \Lambda^{d-3} \Theta(d-3) - \left( \Delta_{SO(4,4)} + 4\right) \cE_{(0,0)} ^{(4,1)}  \delta_{d,4} \log\Lambda
\\ \nonumber&
 - \frac{\Lambda^{\frac{d}{2}-2}}{\frac{d}{2}-2}  \left( \Delta_{SO(d,d)} + d(d-3)\right) \cE_{(0,0)} ^{(d,1)} \Theta(d-4) \Bigg]\,.
\end{align}
Thanks to~\eqref{delsor4l1} the last term in the second line vanishes and the last line is equal to $-\Lambda^{\frac{d}{2}-2}  d\, \cE_{(0,0)} ^{(d,1)} \Theta(d-4)$.  Thus,
\begin{align}
\Big( \Delta_{SO(d,d)} & + d(d-3) \Big) \, \cE_{(1,0)} ^{(d,2)} = \lim_{\Lambda\to\infty} \Bigg[ \pi \int_{\cF_2^\Lambda} \!\de\mu_2\, \Delta_{Sp(4)} \Gamma_{d,d,2}(\Omega)
\nonumber \\ \label{121b} &
- 4\pi d\, \cI_d\, \Lambda^{d-3} \Theta(d-3) 
-\Lambda^{\frac{d}{2}-2}  d\, \cE_{(0,0)} ^{(d,1)} \Theta(d-4) \Bigg]\,.
\end{align}
The contribution of the first line localizes at the boundary of ${\cF_2^\Lambda}$. Decomposing ${\cF_2^\Lambda}$ into $\cF_2^{0} \cup \cF_2^{I} \cup \cF_2^{II}$ as in \eqref{defregIII}, the boundary $t=\Lambda$ of region I corresponds to the minimal separating degeneration, while the boundary 
$V=\tau_2/\Lambda$ of region II corresponds to the maximal separating degeneration. Contributions from the  boundary $\rho_2=\Lambda_1$ of region I, and  $V=1/(\tau_2 \Lambda_1)$ of region II, cancel when the results are expressed in terms of renormalized couplings as we saw in Section~\eqref{sect:rend4r4}.

To analyze the boundary contribution from either region, we note that the Laplacian 
$\Delta_{Sp(4)}$ in the coordinates adapted to each region decomposes into
\begin{align}
\label{eq:dsp4appI}
I\ : \quad &  \Delta_{Sp(4)} =  t^2 \partial_t^2 - t \partial_t + \rho_2^2 \left[\partial^2_{\rho_1} + \partial^2_{\rho_2}\right] +  \ldots\\
\label{eq:dsp4appII}
II\  : \quad &     \Delta_{Sp(4)} =  \frac{1}{2} V^2\partial^2_V + 2 V \partial_V  + \frac{\tau_2^2}{2} \left[\partial^2_{\tau_1} + \partial^2_{\tau_2}\right] + \ldots 
\end{align}
where the omitted terms vanish when acting on functions of $(t,\rho)$ and $(V,\tau)$, respectively. It follows that, for $d\neq 2$,
\be
\begin{split}
\pi \int_{\cF_2^{\Lambda}} \!\de\mu_2\, \Delta_{Sp(4)} \Gamma_{d,d,2}
= ~ &2 \, \left[\frac{1}{t} \partial_t\, t^{d/2}\right]_{t=\Lambda}
\, \cE^{(d,1)}_{(0,0)} 
 -4 \pi\, \int_{\cF_1/\IZ_2}\!\!\!\! \de\mu_1(\tau) \, \left[V^4 \partial_V V^{-d}\right]_{V=\tau_2/\Lambda}
\\
=~ & {d}\, \Lambda^{\tfrac{d}{2}-2}\, \cE_{(0,0)} ^{(d,1)} +
4\pi d\, \cI_d\,\Lambda^{d-3}  \ .
\end{split}
\ee
For $d=2$, the last term is replaced by a term proportional to $1/\Lambda$, which is irrelevant in the limit $\Lambda\to\infty$. Comparing with \eqref{121b}, we see that the divergent $\Lambda$-dependent terms cancel so that
\be
\left( \Delta_{SO(d,d)}  + d(d-3) \right)\, \cE_{(1,0)} ^{(d,2)}   =  24\zeta(2)\, \delta_{d,3}
+4 \cE_{(0,0)}^{(d,1)} \delta_{d,4}\ .
\label{delsod4r42}
\ee
This establishes Eq. \eqref{delsod4r4}, with the correct value of the anomalous terms, and makes it clear that the anomalous terms for $d=3$ and $d=4$ originate from primitive divergences and one-loop subdivergences, respectively.

\subsection{Two-loop \texorpdfstring{$D^6 \cR^4$}~\label{sec_laplaced6r4}}

The analysis of~\eqref{delsod6r4} follows similar steps starting from the definition of the renormalized coupling~\eqref{dsixrfour2R}. Using~\eqref{DeltaGamma} to turn the action of 
$\Delta_{SO(d,d)}$ on $\Gamma_{d,d,h}$ into the action of $\Delta_{Sp(4)}$ on the same, we find
\be
\label{d6r4t1}
\begin{split}
\left( \Delta_{SO(d,d)}  - (d+2)(5-d) \right)\, \cE_{(0,1)} ^{(d,2)}  = &
\lim_{\Lambda\to\infty} \left[ 2 \pi \int_{\cF_2^\Lambda} \!\de\mu_2\, \varphi(\Omega) (\Delta_{Sp(4)} -5) \Gamma_{d,d,2}(\Omega) \right. \\
& \left.
-  \left( \Delta_{SO(d,d)} -(d+2)(5-d)\right)\, e_{(0,1)} ^{(d,2)}(\Lambda)
\right]\ .
\end{split}
\ee
Integrating by parts and using the key property the Kawazumi-Zhang invariant \cite{D'Hoker:2014gfa}
\be
 \left( \Delta_{Sp(4)} -5 \right) \, \varphi=0\ ,
 \ee
valid away from the separating degeneration, we get contributions from i) the boundary $t=\Lambda$ of region I, corresponding to the minimal non-separating degeneration,
ii) the boundary $V=\tau_2/\Lambda$ of region II, corresponding to the maximal non-separating
degeneration and iii) from the boundary $v=0$ of region 0, corresponding to the separating degeneration:
\be
\label{delta3}
2 \pi \int_{\cF_2^\Lambda} \!\de\mu_2\, \varphi(\Omega) (\Delta_{Sp(4)} -5) \Gamma_{d,d,2}(\Omega) 
= \delta_{I} + \delta_{II} + \delta_{S}\ .
\ee
The contributions to $\delta_I$ originate from the $\cO(t)$ and $\cO(1/t)$ terms in \eqref{kzmindeg2},
\be
\label{deltai}
\begin{split}
\delta_I = & \frac{2\pi}{3} \left(\tfrac{d}{2}-1\right) \Lambda^{\frac{d}{2}-1}\, \cE^{(d,1)}_{(0,0)}\, \Theta(d-2)
 \\
+ & \frac{5}{\pi} \left(\tfrac{d}{2}+1\right) \Lambda^{\frac{d}{2}-3}\, \cE^{(d,1)}_{(1,0)} \, \Theta(d-6) + \frac{20}{\pi}  \cE^{(6,1)}_{(1,0)}\, \delta_{d,6}\ .
\end{split}
\ee
The contributions to $\delta_{II}$ originate from the $\cO(1/V)$ and $\cO(V^2)$ terms in \eqref{KZregII},
\be
\label{deltaii}
\begin{split}
\delta_{II} =& 
\frac{4\pi^2 \cI'_d }{3} (d-1)  
 \Lambda^{d-2} \Theta(d-2) + \frac{4 \pi^2}{3} \delta_{d,2} \log\Lambda 
 \\  &  +
 16\pi^2 \Lambda^{d-2} \left(\frac{1}{(d-2)^2} - \frac{{\cal I}'_d}{d-2} \right)
 \Theta(d-2) 
- \frac{4\pi^2}{3}\log\Lambda \, \delta_{d,2}
\\ & + 
\frac{10 \zeta(3)}{\pi}  (d+2) \,\cI_{d-2}\,
 \Lambda^{d-5} \, \Theta(d-5) 
+ \frac{70 \zeta(3)}{3} \delta_{d,5} 
\,,
\end{split}
\ee
where $\cI'_d$ was defined in~\eqref{defIdp}. Finally, the contribution to $\delta_S$ originates from the  logarithmic singularity~\eqref{kzsepdeg} of the Kawazumi-Zhang invariant,
\be
\label{deltaS}
\delta_{S} = - \left[ \cE^{(d,1)}_{(0,0)} + 2\pi 
\frac{\Lambda^{\frac{d}{2}-1}}{\frac{d}{2}-1} \Theta(d-2)\right]^2 \! - 4\pi \log\Lambda \, \delta_{d,2} \, \cE^{(2,1)}_{(0,0)} - 8\pi^2 (\log\Lambda)^2 \delta_{d,2} \ .
\ee

It is worth stressing that the contribution of the $\cO(1/V)$ term in $\varphi(\Omega)$ to $\delta_{II}$, displayed on the first two lines of  \eqref{deltaii} involves two distinct contributions. The first line, proportional to $\cI'_d$, arises upon integrating by parts the term $\frac{1}{2} V^2\partial^2_V + 2 V \partial_V$   inside the Laplacian \eqref{eq:dsp4appII}, and retaining the boundary term at $V=\tau_2/\Lambda$.  The  second line arises instead by integrating by parts the term $\frac{1}{2} \tau_2^2(\partial_{\tau_1}^2+\partial_{\tau_2}^2)$   in \eqref{eq:dsp4appII}, and retaining the boundary term at $\tau_2=\Lambda V$. To see this, we rewrite  the integration domain~\eqref{fd2l} so as to integrate first on $\tau_2$ and then on $V$ and $\tau_1$,
\begin{equation}
  \label{Vtaurange}
 \cF_2^{\Lambda,II}=\{    0\leq \tau_1 \leq \tfrac{1}{2}\,,~|\tau|^2 \geq 1 \,,~
  \tau_2 \leq {\rm min}\left(\tfrac{1}{V \Lambda_1},\Lambda V \right)\,,~~
    \tfrac{\sqrt{1-\tau_1^2}}{\Lambda} < V < \tfrac{1}{\Lambda_1 \sqrt{1-\tau_1^2}}\}\ .
\end{equation}
The integral over $\tau_2$ reduces to a boundary term at $\tau_2=\Lambda V$
whenever $V<1/\sqrt{\Lambda \Lambda_1}$, 
\be
\begin{aligned}
-\frac{4\pi^2}{3} \int_0^{1/2} \de\tau_1 & \int_{\sqrt{1-\tau_1^2}/\Lambda}^{1/\sqrt{\Lambda \Lambda_1}}
2 \de V\, V^{1-d}\, \partial_{\tau_2} A(\tau_1,\tau_2)\Big|_{\tau_2=\Lambda V} 
= \\  
& -\frac{8\pi^2}{3} \Lambda^{d-2} 
\int_{\cF_1/\IZ_2} \de \tau_1 \de \tau_2\, \tau_2^{1-d} \partial_{\tau_2} A(\tau_1,\tau_2) 
\end{aligned}
\ee
where in the second line we renamed $\Lambda V=\tau_2$ and dropped again
$\Lambda_1$-dependent terms. Integrating by parts and using \eqref{defIdp}, this is
\be
-\frac{4\pi^2}{3}(d-1) \Lambda^{d-2} \cI'_d +\frac{8\pi^2}{3} \Lambda^{d-2}\, \int_0^{1/2} \de\tau_1\, (1-\tau_1^2)^{\frac{1-d}{2}} A\left(\tau_1,\sqrt{1-\tau_1^2}\right)\ .
\ee
The integral can be expressed in terms of $\cI'_d$ using \eqref{Idbound2}, leading to
\be
-\frac{4\pi^2}{3}\left[
\left( d-1  - \tfrac{(d+2)(d-5)}{d-2} \right) \cI'_d \,- \frac{12}{(d-2)^2}\right]  \Lambda^{d-2}\ .
\ee
This  explains the second line of \eqref{deltaii}. 

Finally, the second line of \eqref{d6r4t1}, which we shall denote 
by $\delta_e$, evaluates to
\begin{align}
\delta_e = & -\left[ \Delta_{SO(d,d)} 
+\, (d+2)(d-5) \right] \,e_{(0,1)} ^{(d,2)} \nonumber
\\=&
\frac{\pi d}{3}  \Lambda^{\tfrac{d}{2}-1}\,  \cE_{(0,0)} ^{(d,1)}\, \Theta(d-2) - \frac{4\pi^2}{3} \log\Lambda \, \delta_{d,2} 
\nonumber\\
 & - (d+2)(d-5)\left[ \frac{\pi}{3} \frac{\Lambda^{\tfrac{d}{2}-1}}{\frac{d}{2}-1}  \cE_{(0,0)} ^{(d,1)}\, \Theta(d-2) 
 +\frac{\pi}{3} \log\Lambda \, \delta_{d,2} \,\cE_{(0,0)} ^{(2,1)} \right]
\nonumber \\
 & - \frac{10 \zeta(3)}{\pi} (d+2)\cI_{d-2}\, \Lambda^{d-5}  \Theta(d-5) 
 +\left(\frac12 (d+2)(d-4) -(d+2) (d-5)\right)\, 
\nonumber \\
 &\quad\quad \times 
 \left[ \frac{5}{2\pi} \frac{\Lambda^{\frac{d}{2}-3}}{\frac{d}{2}-3} \cE^{(d,1)}_{(1,0)} \Theta(d-6) + \frac{5}{2\pi} \log\Lambda \cE^{(6,1)}_{(1,0)} \delta_{d,6}\right]
\label{te2} \\ 
& - \frac{4\pi^2}{3} \frac{(d+2) (d-5)}{d-2}\, 
 \cI'_d \, \Lambda^{d-2}\,\Theta(d-2) 
+ 8\pi^2 (\log\Lambda)^2 \, \delta_{d,2} + 4 \pi^2 \log\Lambda \, \delta_{d,2} \ .
\nonumber
\end{align}
Here we used the one-loop results~\eqref{delsol1}, and refrained from simplifying some terms 
in order to make it easier to trace their origin either from the constant term or the action of the Laplacian.
Summing~\eqref{te2} and \eqref{delta3}, all $\Lambda$-dependent terms cancel, and we find 
the differential equation for the renormalized $D^6 \cR^4$ coupling,
\be
\label{delsod6r4bis}
\left( \Delta_{SO(d,d)}  - (d+2)(5-d) \right)\, \cE_{(0,1)} ^{(d,2)}  =
- \left ( \cE_{(0,0)} ^{(d,1)} \right ) ^2
+\frac{70}{3}\zeta(3) \delta_{d,5} + \frac{20}{\pi}  \cE_{(1,0)}^{(6,1)}  \delta_{d,6} 
\ee
This establishes \eqref{delsod6r4}, with the correct coefficients for the anomalous terms in $d=5$ and $d=6$, originating from the primitive two-loop divergences and one-loop subdivergences, respectively.

It is worth noting however that no anomalous terms appears in $d=2$ within our renormalization scheme. The reason is that unlike the anomalous terms in $d=5$ and $d=6$, which are annihilated by the operator $ \Delta_{SO(d,d)}  - (d+2)(5-d)$, the anomalous term
$\tfrac{\pi}{3} \cE^{(2,1)}_{(0,0)}+\tfrac{7\pi^2}{18}$ is not, and can be removed by shifting 
$\cE_{(0,1)}^{(d,2)}$ by a suitable multiple of $\cE_{(0,0)}^{(d,1)}$ and a suitable constant.
At the level of the non-perturbative $D^6\cR^4$ coupling, this amounts to a shift of 
$\cE_{(0,1)}^{(2)}$
by a multiple of $\cE_{(0,0)}^{(2)}$ and an additive constant, and must be accompanied by a shift of $\cE_{(0,1)}^{(2,1)}$ by a constant and a shift of the non-analytic part of $\cE_{(0,1)}^{(2)}$ by a multiple of $\log g_8$. The  anomalous term  on the r.h.s. of \eqref{delsod6r4} for $d=2$ was dictated by a choice of renormalization scheme such that no anomalous term appears in the U-duality invariant differential equation for $\cE_{(0,1)}^{(2)}$, while the current
scheme ensures that no anomalous term appears in the T-duality invariant differential equation for $\cE_{(0,1)}^{(2,2)}$.

\section{Discussion \label{sec:discussion}}

The main point of this paper is to show how to explicitly derive the couplings in the low-energy superstring effective action starting from string amplitudes. We focused on the $\cR^4$, $D^4\cR^4$ and $D^6\cR^4$ terms in the effective action of toroidally compactified type II superstring. These
terms can be obtained from the four graviton amplitude by expanding up to $\cO(p^{14})$ in momenta. While the string amplitude is both UV and IR finite for generic values of the graviton momenta, it is convenient to study the low-energy limit by separating the contributions involving the propagation of massless states from the purely stringy contributions. This also provides a natural splitting between the local and the non-local part of the effective action.

This can be efficiently done by introducing appropriate IR cutoffs on the period matrix of the complex structure of the string worldsheet, which can be interpreted as UV cutoffs on the Schwinger parameters of the corresponding field theory diagrams. At two loops and higher some care is required because the string worldsheet can degenerate into a worldsheet of lower genus decorated by propagators of  massless supergravity states. The resulting IR divergences need to be subtracted 
in order to define the stringy contribution to the effective action.
At the two-loop level, this is summarised in Figure \ref{figreg}, where region I contains the 1-loop subdivergences while region II contains the primitive divergence.

Having defined the local terms of the superstring effective action in this fashion, we have shown that they satisfy Laplace-type differential equations with respect to the moduli of the internal torus, and found perfect agreement with predictions from U-duality. This supports the existing conjectures for the exact non-perturbative $D^4\cR^4$ and $D^6 \cR^4$ couplings. Further support could be gained by studying the behavior of the two-loop couplings in the limit where the radius of one circle inside $T^d$ is taken to be much larger than the string scale, and reproducing the pattern of decompactification limits found in \cite{Pioline:2015yea}. 

One reason to focus on the two-loop $D^6\cR^4$ amplitude is the conjecture made in \cite{Pioline:2015yea} that $\cE^{(d,2)}_{(0,1)}$ for $d=5$ provides the exact $D^6\cR^4$ coupling in M-theory compactified on $T^5$ -- largely thanks to the fact that the T-duality group $SO(d,d,\IZ)$ coincides with the U-duality group $E_{d'+1}(\IZ)$ for $d=5, d'=4$. In order to extract the non-perturbative corrections predicted by this conjecture, we have to study the limit in which $T^5$ degenerates into $T^4\times S^1$, which is an instance of the decompactification limit mentioned above. In this work, we have 
laid the ground for this study, by giving a mathematically precise definition of the renormalized coupling
$\cE^{(5,2)}_{(0,1)}$. 

Clearly, it is also desirable to extend this analysis to the three-loop contribution to the $D^6 \cR^4$. The latter is proportional to the modular integral of the lattice partition function $\Gamma_{d,d,3}$ over the Siegel upper half-plane of degree three \cite{Gomez:2013sla,D'Hoker:2014gfa}, but the latter diverges when $d\geq 4$ while one-loop subdivergences and two-loop divergences set in when $d=5$ and $d=6$, respectively. We plan to investigate the differential equation satisfied by $\cE^{(d,3)}_{(0,1)}$ and its decompactification limits in future work.

Finally, it would be very interesting to extend the methods of this work to a more general class of two-loop amplitudes beyond the simple BPS-saturated amplitudes considered here, such as $D^8\cR^4$ amplitudes in type II theories, or two-loop amplitudes in heterotic string theory.  A particularly interesting example is the $D^2 H^4$ amplitude in type IIB compactified on $K_3$, which is shown to 
satisfy a  differential equation similar to \eqref{delsod6r4}, and conjectured 
to be given non-perturbatively by a two-loop heterotic modular integral where the integrand has a pole on the separating divisor \cite{Lin:2015dsa}.

\bigskip

\acknowledgments

We would like to thank E.~d'Hoker and M.~Green for discussions and collaboration on \cite{D'Hoker:2014gfa}, which paved the way for the present work, and to G.~Bossard and  I.~Florakis for useful discussions. This research is partially supported by STFC (Grant ST/L000415/1, {\it String theory, gauge theory \& duality}).

\providecommand{\href}[2]{#2}\begingroup\raggedright\endgroup

\end{document}